\documentclass[aps,twocolumn,preprintnumbers,amsmath,amssymb]{revtex4}

\usepackage{graphicx}
\usepackage{dcolumn}
\usepackage{bm}
\newcommand{\be}[1]{\begin{equation}\label{#1}}
\newcommand{\ee}{\end{equation}}

\begin{document}

\title {Fatigueless response of spider draglines in cyclic torsion facilitated by reversible molecular deformation } 

\author{Bhupesh Kumar and Kamal P. Singh}

\affiliation{Department of Physical Sciences, Indian Institute of Science Education and Research 
 Mohali, SAS Nagar, Sector 81, Manauli 140306, India.}

%\date{3 Oct. 2014}

\begin{abstract}
We demonstrate that spider draglines exhibit a fatigueless response in extreme cyclic torsion up to its breaking limit.     
The well defined Raman bands at $1095$ and $1245~cm^{-1}$ 
shifted linearly towards lower wavenumbers versus increasing twist in both clockwise and counter-clockwise directions.   
Under thousands of continuous loading cycles of twist strain approaching its breaking limit, all the Raman bands were preserved
and the characteristic Raman peak shifts were found to be reversible. Besides, nanoscale surface profile of the worked silk appeared as good as the pristine silk. This unique fatigueless twist response of draglines, facilitated by reversible deformation of protein molecules, 
could find applications in durable miniatured devices.

\end{abstract}

\maketitle

The spider dragline silk is a wonderful material having a unique combination of high
tensile strength and elasticity \cite{Rev_FV, Rev_DK, Rev_DP, Rev_TS, Rev_JM, Rev_MB}. 
Due to its remarkable physical properties and biocompatibility, many application were 
demonstrated such as its use as biological optical fibers \cite{Huby} and in uniquely 
packed guitar strings \cite{violen}. There is a great interest in understanding 
structure-property function relationship of silk subjected to variety of deformations. 

The macroscopic tensile properties of silk and underlying molecular mechanisms
are well established using variety of experimental and theoretical techniques \cite{Rev_FV, Rev_DK, Rev_DP, Rev_TS, Rev_JM, Rev_MB}.
It is now well known that its high tensile toughness originated due to cooperative
action of hydrogen bonded $\beta$-sheet nanocrystals \cite{Atomic-1, MJB-1} while the amorphous regions provide 
elasticity \cite{Rev_FV, Atomic-1, Backbone_PNAS}. 
The Raman microspectroscopy has also been used to study various stress induced molecular deformations
\cite{Vollarath_Raman, Raman-FV, Lefevre, Young1, Young2}.
However, response of silk to other forms of deformation such as torsion has attracted much less attention.

Previous measurements on macroscale torsion response reported its shear modulus \cite{FKO}, 
small angle self-recovery of twist \cite{Olivier} and a torsional superelasticity, i.e., its 
ability to reversibly withstand great torsional strain without rupture \cite{KPS}. 
Surprisingly, molecular deformation of silk under torsion has remained unknown. 
Furthermore, it is also fundamentally important to characterize fatigue response of spider draglines,
 for example in extreme cyclic torsion up to breaking limit.

In this letter, we report first molecular deformation study and fatigue response of spider draglines 
subjected to thousands of torsional loading cycles.  
By twisting the draglines in controlled fashion and employing Raman micro-spectroscopy,    
we show that some well defined Raman peaks shift 
linearly towards lower wavenumbers with increasing twist strain for both clockwise (CW) and counter clockwise (CCW) directions.
Remarkably, we report a fatigueless response of silk whereby after thousand loading cycles of twist strain  
approaching its breaking limit, the Raman shifts were reversible. Besides, the 
nanoscale profile of the worked silk was found to be same as that of the pristine silk.

The dragline silks were obtained from the female spider (Areneous Neoscona)
harnessed in laboratory at $25^\circ C$ and $50~\%$ humidity \cite{KPS}. The spider was made to jump from a small
plastic rod following which it suspended itself from the dragline silk. The typical diameter (D)
of the silk filaments was around $2-4~\mu m$ as measured from scanning electron microscope (SEM) images (Fig.~1). 
The silk samples were carefully mounted on the torsion setup (Fig.~1(a)) without
any stretching and twisting. The draglines were then twisted by prescribed amount 
from one end using a calibrated motor rotating at 35~Hz.
The direction of twist (CW or CCW) was controlled by reversing the
driving voltage ($\pm 8~V$) of a square pulse from a digital signal generator. The number of given turns, $\tilde{N}=N/L$ per unit length of the silk, was measured with $\pm 1\%$ accuracy by recording a video and revolutions of the motor.  

The Raman microspectroscopy has proven to be a successful
technique to access molecular deformation of silk monofilaments \cite{Vollarath_Raman, Raman-FV, Lefevre} and other polymer fibers 
\cite{Young1, Young2}. Since the Raman 
spectra of the silk is known to sensitively depend on tensile strain, we took special care to transfer the twisted silk from our torsion setup onto the slide without any stretching.
The Raman spectra of the silk were recorded using a Raman 
spectrometer (Renishaw inVia) having a resolution of $2~cm^{-1}$. 
A 5-10 mW He-Ne laser at $633~nm$ wavelength was  
focused with 100X Leica objective on the monofilament of dragline. 
Each Raman spectra was averaged over 5 scans at various location on the silk with $100~s$ exposure
for each spectrum. No damage of our sample after the Raman spectra was observed.

Three different kinds of experiments were performed. (i) Raman spectra
of various draglines were recorded versus increasing twist up to the
breaking limit, (ii) effect of CW and CCW twist direction was studied, and 
(iii) fatigue response of draglines in cyclic torsion was characterized using Raman and SEM imaging.    

%%%%
 \begin{figure}[t]   %  Figure 1
    \includegraphics[width=.99\columnwidth]{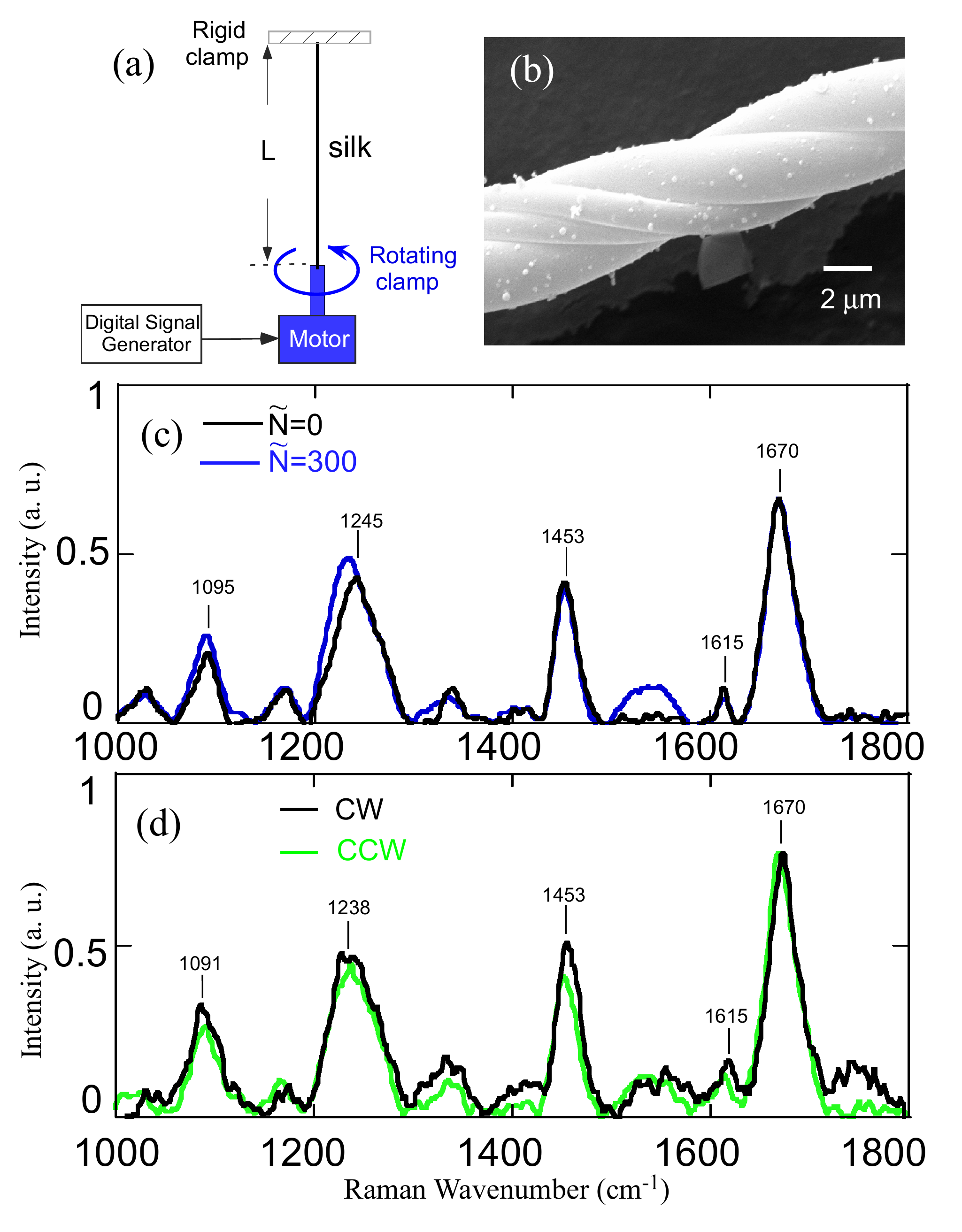}
\caption{ (a) Schematics of the torsion set-up. Typical length of the silk sample was $L=5~cm$. 
(b) A SEM image of a twisted Au coated dragline silk.
 (c) Raman spectra for an untwisted dragline silk (black) and a twisted silk (blue). (d) Comparison of 
 Raman spectra of CW and CCW twists for identical turns per cm, $\tilde{N}=300$.}
\label{fig:fig1}
\end{figure}
%%%%

%%%%%---  Figure 2 -----------
\begin{figure}[t]
  \begin{center}
    \includegraphics[width=.99
    \columnwidth]{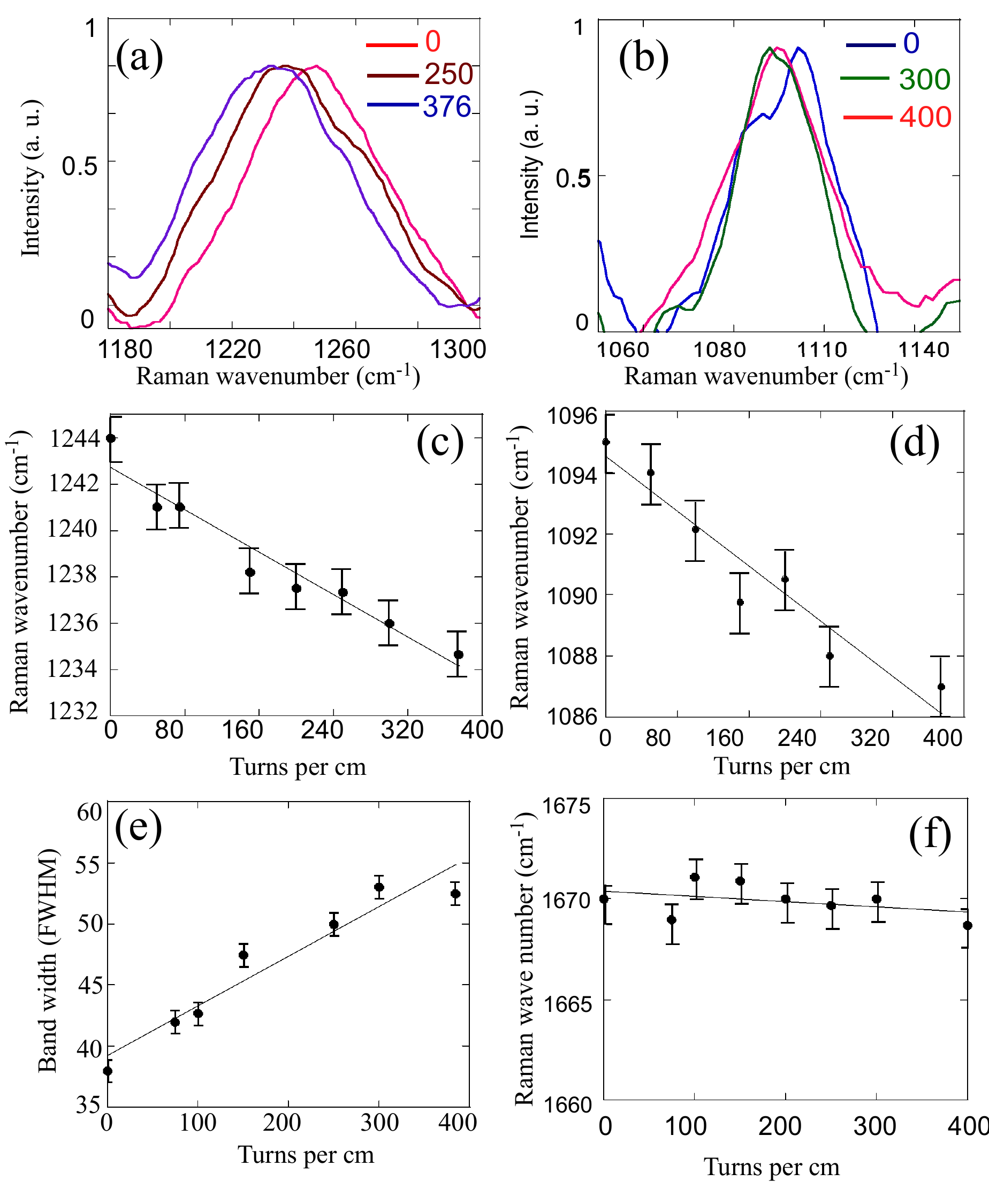}
     \caption{ The Raman bands at (a) 1245 $cm^{-1}$ 
     and (b) 1095 $cm^{-1}$ versus increasing $\tilde{N}$ turns per cm as labeled in the plots. 
     (c), (d) Average peak positions of these bands versus $\tilde{N}$. Each data point is averaged over five
     measurements at different location on the twisted fiber. Solid lines are linear fits. Error bar represents 
     instrumental resolution.
(e) FWHM of 1245~$cm^{-1}$ band versus increasing turns per cm, (f) Peak position of 1670~$cm^{-1}$ band versus $\tilde{N}$.}
\label{fig:fig2}
   \end{center}
\end{figure}
%%%%

First, we compared Raman spectra of a native untwisted dragline with a twisted one in Fig.~1(c).  
Five prominent well assigned Raman bands, including amide I and III regions of the silk proteins, were 
observed at 1095, 1245, 1453, 1615, 1670 $cm^{-1}$. 
We observed that amide I peak at 1670~$cm^{-1}$ due to C=O stretch vibration in $\beta$-sheet and the 
tyrosine side chain at 1615~$cm^{-1}$ were mostly unaffected by twist. This was probably due to 
the fact that the carbonyl group and the side chain are oriented perpendicular the 
fiber axis \cite{Raman-FV, Lefevre_Review, Lefevre, JPhysChem}.
Also, the 1453~$cm^{-1}$ peak, assigned to C-H bending mode which is reminiscent of orientation of Gly and Ala residue in
$\beta$-sheet \cite{Lefevre_Review, Maiti}, was hardly affected by twist strain.
Notably, 1245~$cm^{-1}$ band assigned to C-N stretch vibration in $\beta$-sheet and 
1095~$cm^{-1}$ band due to C-C vibration in random conformation, both parallel to the polypeptide chain, exhibited significant redshift.  
The C-N stretch vibration was particularly sensitivity to
alteration in internal rotation angle and bond length due to its partial double-bond character and 
their orientation parallel to the fiber axis \cite{Lefevre_Review, Maiti}.
 Therefore, the sensitivity of 1095 and 1245 $cm^{-1}$ peaks to twist will be used to follow resulting molecular deformation. 

We used many samples obtained from the spider under identical conditions and 
twisted them one by one by $\tilde{N}=0$ to 400 turns/cm in steps of 50 turns/cm. 
To obtain their Raman spectra, the twisted samples were carefully transferred to the glass slides without introducing
any stretching or twisting. 
As shown in Fig~2(a)-(d), the peak position of the Raman bands at 1095 and 1245~$cm^{-1}$ shifted linearly towards lower wavenumbers with increasing turns $\tilde{N}$. The peak positions were computed 
by averaging over 5 spectra taken at random locations on each sample. The rates of peak shift for 1245~$cm^{-1}$ and 1095~$cm^{-1}$ peaks were -2.2~$cm^{-1}$ and -2~$cm^{-1}$ per 100 turn/cm, respectively. These Raman shifts indicated that the 
macroscopic twist is directly transferred to
the bonds in the silk molecules which alters its bond length and internal rotation angles. 
A linear dependence of Raman shift with $\tilde{N}$ suggests a uniform stress distribution in the fiber. This means that the twist induced bond deformation is higher in the softer amorphous region compared to the rigid crystalline ones. A similar linear Raman shift was also observed under tensile loading condition for spider and silkworm silks \cite{Vollarath_Raman, Young1}. We also estimated twist induced  broadening of $1245~cm^{-1}$ band. Its full width at half maxima (FWHM) was found to be around 4.0~$cm^{-1}$ per 100 twist/cm (Fig.~2(e)),
which indicated an increase in heterogeneity due to local stress distribution in the silk molecules
\cite{Vollarath_Raman, Young1, Young2}. A more detailed analysis of all the 
aspects of the spectra merits further study.
 
We also compared the response of the silk for CW and CCW loading. Extracting a long
dragline, we defined a reference fiber axis and made two samples ($L=5~cm$) from it. With respect to the axis direction, one sample was twisted in CW direction and the other one in CCW by identical number of turns $\tilde{N}=300 \pm 5$ per cm. The obtained Raman spectra for both the samples are compared in Fig.~1(c). Using 10 spectra per sample on four different samples, we observed that for identical $\tilde{N}$ the Raman spectra superimposed on each other within experimental precision. The Raman peaks at 1095 and 1245 $cm^{-1}$ shifted towards lower wavenumbers for both the CW and CCW turns. 

\begin{figure}[t]    %--------- Figure 3  ------------
\includegraphics[width=.99\columnwidth]{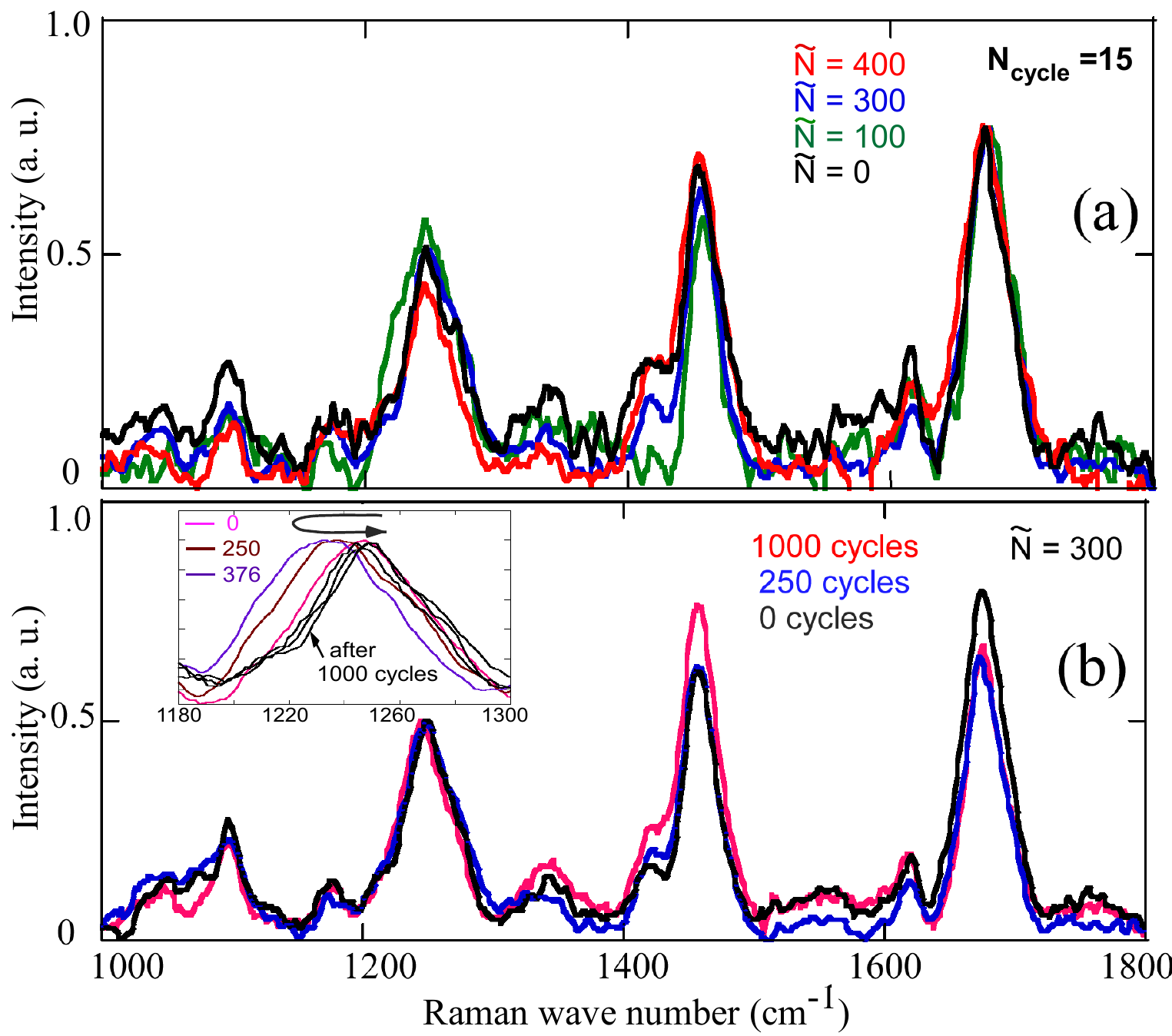}
\caption{ Raman spectra of draglines for, (a) different $\tilde{N}$= 0, 100, 300, 400 turns per cm
and fixed $N_{cycle}=15$, (b) increasing twist cycles $N_{cycle}$= 0, 250, 1000 for fixed $\tilde{N}=300$ turns per cm. 
Inset: recovery of $1245~cm^{-1}$ band following 1000 loading cycles, $\tilde{N}$ values are labeled.
}
\label{fig:fig3}
\end{figure}

What is the fatigue response of fine draglines in cyclic torsion?  
Answering this question is important for understanding dynamical response of the silk as well as 
for its potential applications. 
We subjected draglines to cyclic twist
by driving the motor with a square pulse $\pm8~V$ with $76~s$ time period for each cycle. In one loading cycle,  
the silk was twisted in CW direction by $\tilde{N}$ turns per cm and then untwisted to zero turn.   
Two types of fatigue tests were done. First, the number of twist cycles were fixed to $N_{cycle}=15$ and 
varied $\tilde{N}=0$ to $400$ turns per cm, which is close to its breaking limit.  
The corresponding Raman spectra were almost identical as shown in Fig.~3(a). 
Second, we fixed $\tilde{N}=300$ and increased twist cycles to $N_{cycle}=1000$ as shown in Fig.~3(b).
Note that the net twist in the worked silk was zero in all these cases.
One can see in Fig.~3 that all the prominent Raman peaks were preserved and their locations were 
statistically identical to the ones for the fresh silk. In particular, the 1245~$cm^{-1}$ band
clearly exhibited reversibility after 1000 cycles as shown in inset of Fig.~3(b). 

\begin{figure}[t]    %--------- Figure 4  ------------
\includegraphics[width=.99\columnwidth]{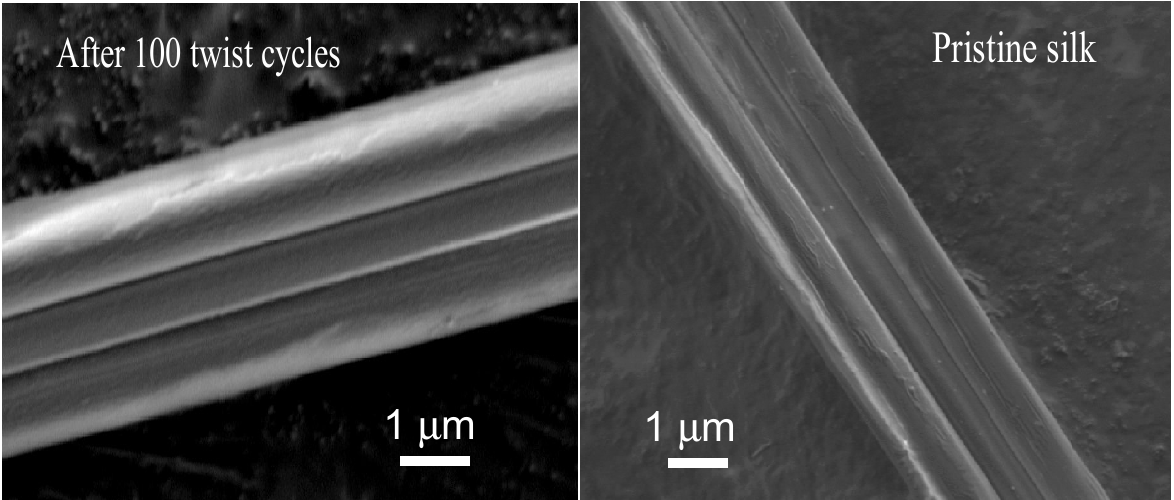}
\caption{ Comparison of the SEM images of dragline silk following $N_{cycle}=100$ loading cycles with $\tilde{N}=300$ (left) with pristine silk (right). }
\label{fig:fig4}
\end{figure}

Furthermore, we compared nano-scale silk-surface after cyclic loading with the pristine silk (Fig.~4). 
Despite large shear strain ($\tilde{N}D \approx 0.01$) in each cycle, the SEM image of the worked silk 
appeared identical to that of the pristine silk. 
Supplementary optical diffraction measurements on the worked silk (not shown) confirmed recovery of its original diameter
within $100~nm$. 
It is worth mentioning that microscale metal (Cu, Au) wires of $20-100~\mu m$ diameter, under much less strain $\tilde{N}D \approx 0.001$, 
exhibit anomalous plasticity and fatigue. In contrast, they fail after few torsion cycles and display cyclic bands and cracks on its surface \cite{WireTorsion}. Clearly, the silk belongs to a unique class of natural microscale fiber that exhibited a fatigueless response
whereby the macro-scale twist is distributed to molecular bonds in the silk proteins that is reversible in cyclic torsion.

In summary, we report a unique fatigueless response of spider draglines in cyclic torsion of
extreme amplitude up to the breaking limit. 
The characteristic Raman bands at 1095 and 1245 $cm^{-1}$ shifted linearly towards lower wavenumbers
with increasing turns per cm for both CW and CCW directions.
The Raman bands corresponding to the carbonyl group and side chains were mostly unaffected by twist strain.
Remarkably, when the silk was subjected to $10^3$ twist cycles up to breaking limit
the Raman bands exhibited reversibility which indicated a reversible molecular deformation in cyclic torsion.
Besides, the nanoscale surface profile of the worked silk appeared identical to the pristine silk. 
The simple strategy evolved by spider draglines could possibly be exploited in man-made polymers resulting in similar properties. 
The unique fatigue-less response, facilitated by reversible molecular deformation, could find 
applications in durable and biocompatible miniatured devices \cite{Biomimetic1}.

% beam, cantilever, 

We are grateful to Bhavin Kansara for his help with the Raman spectrometer. 
We thank Hemaswati, S. Arya, P. Majumdar and Prof. K. S. Viswanathan for Raman spectra.
The central facilities of IISER Mohali and JNU New Delhi is acknowledged. 
KPS acknowledge financial support by DST India through Ramanujan fellowship.

\end{document}